\documentclass[]{article}

\usepackage{graphicx}
\usepackage{url}
\title{An analysis of Principle 1.2 \\ in the new ACM Code Of Ethics}
\author{Christoph Becker\\University of Toronto\\christoph.becker@utoronto.ca\\
\url{https://ischool.utoronto.ca/profile/christoph-becker/}}

\begin{document}

\maketitle

\begin{abstract}

The new ACM Code of Ethics is a much-needed update, but introduced changes to a central principle that have not been discussed widely enough. This commentary aims to contribute to an improvement of the ethical standards we want computing professionals to aspire to by analyzing how changes introduced to Principle 1.2, Avoid Harm, affect the Code as a whole. 

The analysis shows that the principle is now internally inconsistent in structure and externally inconsistent with Principle 2.3. It condones intentional harm too broadly and does not oblige those responsible to seek external justification. The existing Principle 2.3 clearly suggests that Principle 1.2 is unethical. 

As a consequence, the change introduced to Principle 1.2 in the new Code of Ethics nullifies the good intention of the code; counteracts the many good changes introduced in all three drafts; and places the ACM in a dangerous moral position. 

This short paper explains why and recommends concrete actions. 
\end{abstract}

\section{How to Avoid Harm}

There has never been a more important time to raise the bar for ethical standards in computing. This year, the Association of Computing Machinery, the world’s major association of computing professionals and researchers, updated its Code of Ethics \cite{code2018}, previously updated 1992 \cite{code1992}. As part of this update, multiple drafts were released for comments. Many excellent changes were introduced to update and strengthen the code and take into account feedback of the community, and the voluntary leadership, effort and insight of the ACM Code 2018 team deserve praise from the community. 
The Code contains crucial changes to a central principle, however, that weaken the code substantially and fundamentally beyond the point of usefulness. 

This commentary aims to contribute to a further improvement of the standards we want computing professionals to aspire to. Let us not wait another 26 years until adapting the code again. 

The last of three drafts of the Code of Ethics contained a crucial change to the central \textbf{Principle 1.2,'Avoid Harm'}. In draft 3, it read as follows.\footnote{In the final release of the code, paragraph 2 was changed to end with `minimize all harm' following a Twitter thread posted in June 2018 \cite{becker2018tweet} in which I criticized draft 3 (as discussed below). It is unclear what caused this change, since the ACM Ethics team did not respond to the detailed argument I shared with them in July 2018 (a previous version of this document.)}

\begin{quotation}1.2 Avoid harm.\\
	In this document, “harm” means negative consequences to any stakeholder, especially when those consequences are significant and unjust. Examples of harm include unjustified physical or mental injury, unjustified destruction or disclosure of information, and unjustified damage to property, reputation, and the environment. This list is not exhaustive.\\
	Well-intended actions, including those that accomplish assigned duties, may lead to harm. When that harm is unintended, those responsible are obligated to undo or mitigate the harm as much as possible. Avoiding harm begins with careful consideration of potential impacts on all those affected by decisions. When harm is an intentional part of the system, those responsible are obligated to ensure that the harm is ethically justified and to minimize unintended harm.\\
	To minimize the possibility of indirectly harming others, computing professionals should follow generally accepted best practices. Additionally, the consequences of emergent systems and data aggregation should be carefully analyzed. Those involved with pervasive or infrastructure systems should also consider Principle 3.7.\\
	A computing professional has an additional obligation to report any signs of system risks that might result in harm. If leaders do not act to curtail or mitigate such risks, it may be necessary to “blow the whistle” to reduce potential harm. However, capricious or misguided reporting of risks can itself be harmful. Before reporting risks, a computing professional should thoroughly assess all relevant aspects.
\end{quotation}
All this might sound fine, if vague, but a closer look at the text and the recent changes tells a different story. The principle is now internally inconsistent in structure, which opens a loophole, and it is externally inconsistent with Principle 2.3. It omits the need to minimize intentional harm; condones intentional harm very broadly; and does not oblige those responsible to seek external justification. A footnote clarifies that this central change was introduced at this late stage primarily to appeal to the weapons industry. According to the existing Principle 2.3, Principle 1.2 should be considered unethical.

\begin{figure}
	\caption{Changes to Principle 1.2 (new elements in green)}
	\label{changetrack}
	\centering
	\includegraphics[width=11cm]{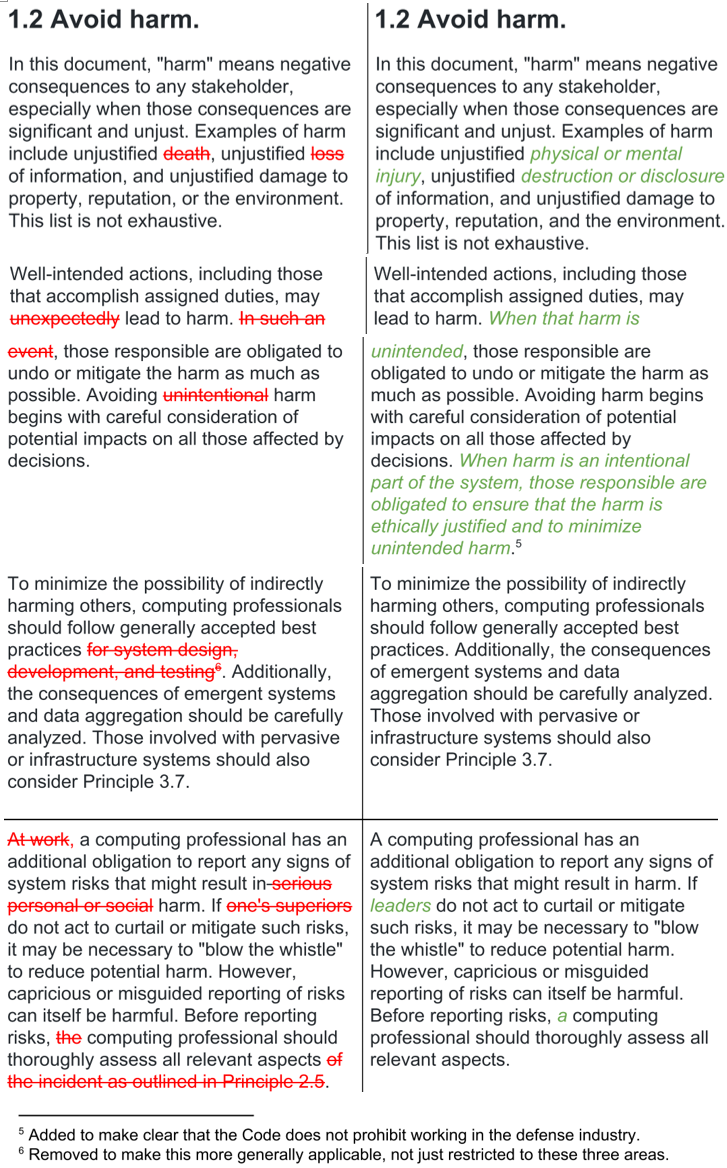}
\end{figure}

As a consequence, the change introduced to Principle 1.2 in the new Code of Ethics nullifies the good intention of the code; counteracts the many good changes introduced in all three drafts; and places the ACM in a cdangerous moral position. This short paper explains why. Figure \ref{changetrack} reproduces the annotated document that highlights the change introduced in draft 3 \cite{acmCOPE}. Below, I explain my concerns one by one. 

\section{So what’s wrong with this?}
I will discuss six major concerns over the current phrasing of Principle 1.2.
\begin{enumerate}
	\item The justification of intentional harm is ill-specified and inconsistent with the proportionality principle.
	\item Principle 2.3 requires us to challenge Principle 1.2 as an unethical rule with inadequate moral basis.
	\item Principle 1.2 misleadingly suggests that internal justification without legitimation is sufficient. 
	\item The definition of harm in Principle 1.2 is inconsistent with the examples provided.
	\item The concept of ‘best practices’ used in Principle 1.2 is vague and inadequate.
	\item The relationship of ACM to the Weapons Industry requires discussion.
\end{enumerate}
Below, I discuss each concern in turn. For additional context, I urge you to review the entire Code.

\subsection{Intentional harm and the ethical principle of proportionality}
Draft 3 of Principle 1.2 introduced a distinction between intentional and unintentional harm, designed to expand the Code's applicability to the military domain. The idea that harm can be easily divided into intentional and unintentional harm and then `divided and conquered' may be appealing, but does it have an adequate moral basis? How actionable is this distinction? 

The ethical principle of \textbf{double effect} states that `Applications of double effect always presuppose that some kind of proportionality condition has been satisfied. Traditional formulations of the \textbf{proportionality condition} require that the value of promoting the good end outweigh the disvalue of the harmful side effect' \cite{mcintyre_doctrine_2014}. The new draft principle introduces distinctions along the lines of double effects, but does not address proportionality.
Principle 1.2 in draft 3 required those responsible to ‘minimize unintentional harm’, but did not oblige them to also minimize intentional harm. Consider the effect on three example scenarios.
\begin{enumerate}
	\item 	Consider a team designing a bomb to end a war and its suffering. Imagine they face a choice between a bomb that would neutralize electronic equipment and a hydrogen bomb. According to the Principle above, the responsible agents would have to minimize unintentional harm, but not intentional harm. This would make the hydrogen bomb the more appealing solution. Of course, other principles, such as 2.3, would suggest otherwise, but how sensible is any principle that does not require us to minimize \textit{all} harm?
	
	\item An example closer to ACM members’ research is the ethical dilemmas posed by self-driving cars and potential harm: As self-driving cars capture the popular imagination, an ongoing conversation and debate focuses on how decisions to trade off on harm are adjudicated. It is heartening to see this discussion draw on long standing conversations in ethics around intention and harm, primarily through the trolley problem, (though perhaps too often \cite{bogost_enough_2018}). These conversations point to important distinctions between intentions and outcomes, the role of proportionality in justifying harm caused through the harm avoided and the problematic nature of utilitarian measures of harm. 
	
	\item More commonplace then: In the process of discussing this document, I and my colleagues discussed the action of emailing all members of a dormant community email list we maintain to alert them to the danger this change poses. This involves minor harm - the annoyance of yet another unsolicited email - for a greater public good - the wider discussion of an ethics code. It may also involve unintentional harm of some minor kind, which we are unable to anticipate. These would be potential side-effects such as a person’s possible irritation about a standpoint they do not share, or the carbon footprint of emails.
	The draft Principle 1.2 required us to minimize these unintentional harms - which we cannot feasibly anticipate - but it did not require us to minimize the intentional harm: the annoyance caused by yet another unsolicited email. It seems obvious and commonsense, though, that we should aim to minimize this annoyance, for example by taking care to explain the rationale behind the message.
	
\end{enumerate}

What is the result of the distinction between intentional and uninentational harm? Harm that is unintentional should be minimized, but this category is notoriously fraught. Unintentional harm is often unseen harm, unanticipated harm. Harm that remains invisible cannot possibly be minimized, however. This must not absolve a computing professional from their responsibility to aim to identify it! 

\textbf{Proposed Actions:} Change the text so that all harm must be minimized. Clarify the conceptual framework used to distinguish intentions from unintentional outcomes. Discuss the double effect principle, clarify how it is interpreted, and incorporate proportionality. 

The first of these changes was introduced to the released  version of the code. In the final release, paragraph 2 was changed to end with \textit{`minimize all harm'}. This was a much-needed change, but it does not address definitional clarity or proportionality.

\subsection{The Code itself suggests Principle 1.2 is unethical.}

As the authors of the Code emphasize, the principles of the code should be read in conjunction, as a whole. While Principle 1.2 applies directly to computing professionals and their actions, Principle 2.3 applies to rules such as the Code itself.  It is reproduced below (my highlights in \textbf{bold}): 
\begin{quotation}
2.3 Know, respect, and apply existing rules pertaining to professional work.\\
“Rules” here includes regional, national, and international laws and regulations, as well as any policies and procedures of the organizations to which the professional belongs. Computing professionals must obey these rules unless there is a compelling ethical justification to do otherwise. \textbf{Rules that are judged unethical should be challenged. A rule may be unethical when it has an inadequate moral basis}, it is superseded by another rule,\textbf{ or it causes recognizable harm that could be mitigated through its violation}. A computing professional who decides to violate a rule because it is unethical, or for any other reason, must consider potential consequences and accept responsibility for that action.
\end{quotation}

Principle 2.3 clearly explains that a rule `may be unethical when it has an inadequate moral basis ... or it causes recognizable harm that could be mitigated through its violation.' The code's authors urge us to consider the code holistically, and Principle 2.3 emphasizes: `Rules that are judged unethical should be challenged.' My challenge is based on two key arguments.

First, the moral basis of Principle 1.2 is not substantiated adequately at this point. It now distinguishes between intentional and unintentional harm (without justifying this conceptual distinction clearly), which points to the ethical principle of double effect. Since there is no mentioning of proportionality in Principle 1.2, however, the concept cannot be relied in support the justification of harm in Principle 1.2 against Principle 2.3. The lack of addressing proportionality and the lack of clarifying the code's position with respect to the difficult question of intentions highlight that the moral basis of Principle is highly questionable.

More bluntly and immediately, an examination of the examples above reveals that by violating Principle 1.2, one could `avoid recognizable harm' easily, because Principle 2.3 does not discuss justification of harm nor distinguish intentional from unintentional harm. There is no doubt that Principle 1.2 `causes recognizable harm that could be mitigated through its violation'. This renders it unethical in light of Principle 2.3. As a result, the entire Code of Ethics is internally contradictory. As Owens stated on the ACM ethics discussion board, `if it is to be the requirement that all members adhere to it, the code must be free from controversy and enforceable. Looking through this code, it is neither in its current state.' \cite{owen}
 
\textbf{Proposed Action:} Start a broad and proactive consultation on this subject and initiate one more revision of the Code with a clear timeline, focused on Principle 1.2. Involve other disciplines fully.

\subsection{Justification and legitimation}
When it comes to justifying harm ethically, Principle 1.2 only states “those responsible are obligated to ensure that the harm is ethically justified and to minimize unintended harm.” It does not clarify where legitimacy comes from, and it does not clarify that justification can ultimately not come from those responsible themselves. Consider the example of academic research ethics: Once a minimum threshold of risk is exceeded, a researcher intending to involve participants in research involving humans or animals must seek approval from an Institutional Review Board with legitimated authority on approving the involvement. This justification is based on (1) the principle of proportionality \textit{and} (2) the legitimate external authority of a Research Ethics Board. In any case of whether the threshold is exceeded, the decision lies with the Research Ethics Board.

These principles are so important because the implications of scientific and technological work on society can never be justified with purely technical reasons grounded in the research or the technology development alone. It is no coincidence that the ACM SIGCHI group has an Ethics committee\cite{sigchi-ethics} that perform similar functions in the context of ACM. At no point does the current Principle 1.2 make clear, however, that legitimation must come from somewhere. 

\textbf{Proposed Action: }Consider proportionality and the Precautionary Principle; consider more fully and discuss publicly what other professional associations (medical, genetic, biological, civil engineering, psychology) have done and how the case of ACM relates and compares to their effort; clarify that ethical justification must in many cases involve external legitimation rather than come from those responsible. The resulting Principle 1.2 must spell clearly how and by which principles (such as proportionality) harm can be justified.

\subsection{Unjustified harm}
All examples given for harm include the term `unjustified', but the definition does not, and rightfully so: Harm is a `negative consequences to any stakeholder' in Principle 1.2. The examples are a legacy of prior versions of the code, which did not discuss intentional harm, hence did not need to discuss justification in detail. The subsequent discussion in Principle 1.2 covers justification, however inadequately, but all examples duplicate the idea that harm is only harm if it is not justified. By only listing ‘unjustified’ outcomes as harm, it becomes too easy to argue that a specific negative consequence should not be seen as harm at all because it is justified. But justified by whom according to which standard?

Additionally, the phrase `especially when those consequences are significant and unjust' renders the definitional clause vague: Either something is a negative consequence or not. Which amount of negative consequence would be the threshold at an outcome is no longer harm? Milk is not `especially' milk when it comes from a cow, and harm is no less a negative consequence if the consequence is relatively mild.

\textbf{Proposed Action:} Justification of harm is and must remain distinct from the definition of harm as negative consequence. Remove ‘unjustified’ from all examples. 

\subsection{“Best practices”}
The principle says that “To minimize the possibility of indirectly harming others, computing professionals should follow generally accepted best practices”. We should all be aware that following `generally accepted best practices' is not a high standard. How have any of these practices helped so far to avoid the harm already done? Should we not all strive to do better, to reflect critically, to learn and improve those practice? 

As the current discussions of the role of Machine Learning algorithms in reinforcing gender inequality and racial bias demonstrate in abundance, `best practice' is much too low a bar.

\textbf{Proposed Action:} Computing professionals should critically reflect on their own practices as well as generally accepted best practice and strive to improve both. The Code therefore must reflect a fundamental commitment to critical reflection and constant questioning of what should count as `best practice' in computing.

\subsection{ACM and the weapons industry}
The footnote to the annotated draft reveals that the change introduced to Principle 1.2 was introduced to `make clear that the Code does not prohibit working in the defense industry'. The term is a euphemism sometimes used to denote the arms industry or weapons industry \cite{Arms}. Regardless of whether we see this as an instance of Newspeak\cite{Newspeak} or not, it points to an important question: What position should the code take with respect to human actions that involve intentional harm, such as the military and associated industries? Computing research has a long and difficult historical relationship with the military. The history of the computing industry’s involvement with the military is no secret either\cite{Leslie}. The immediate responses by some of the code’s authors pointed to the need to include military personnel under the code. But note that military personnel is not part of the arms industry!
My personal stance on whether the code should or should not cover the military is not important. What the principle now effectively articulates, however, is a different story.

Regardless of whether we approve or oppose of the idea that the code should apply to military computing personnel, the idea that this should apply to the manufacturing of killing machines raises the question:\textbf{ What values does ACM want to stand for?} Does ACM want to be in the `business of war' \cite{googleletter}? 

Software is not neutral technology and never was, as the history of technology has shown over and over again. Sometimes, this lack of neutrality is more obvious, such as in the case of the military domain. Sometimes, it is less obvious, as in a classic case of civil engineering \cite{winner_artifacts_1980}. Many recent cases illustrate the role of algorithms in reinforcing bias (\cite{algorithms,weapons,eubanks_automating_2018, Amazon}). The recent protest of Google employees against a project with the Pentagon, and a similar protest of Microsoft employees about the company's work with ICE, reminds us that computing professionals can take a stand. Both cases resulted in the company withdrawing from those involvements. Does ACM as an organization want to fall behind the personal standard of those Google and Microsoft employees? Does ACM want to be forced to examine its ties with the military, as the American Psychological Association needed to do when its involvement with torture became public? 
The principles embodied in this Code exemplify and express the values that ACM stands for. They are not neutral, and neither can ACM be `neutral'. In technology design, these values become facts: \begin{quotation} `Values are the facts of the future. Values are not the opposite of facts, subjective desires with no basis in reality. Values express aspects of reality that have not yet been incorporated into the taken for granted technical environment. That environment was shaped by the values that presided over its creation. Technologies are the crystallized expression of those values. New values open up established designs for revision.’\cite{feenberg_ten_2010}
\end{quotation}
\textbf{Proposed Action: }The ACM leadership needs to clarify where ACM stands to enable ACM members to take their own decisions about whether they can identify with the values the organization embodies.

\section{Conclusions: After the Code is before the Code}
The discussion above shows that despite best intentions, the Principle 1.2 is left in need of serious revisions. Principle 1.2 suggests that internal justification without legitimation is sufficient, but it is not. The definition of harm in Principle 1.2 is inconsistent with the examples provided. The justification of intentional harm that it articulates is ill-specified and inconsistent with the proportionality principle, and as a consequence, Principle 2.3 requires us to challenge Principle 1.2 as an unethical rule with inadequate moral basis. In addition, the concept of `best practices' used in Principle 1.2 is vague and inadequate. 

As a result, Principle 1.2 in the new Code of Ethics nullifies the good intention of the code and places ACM in a dangerous moral position. This is exemplified by the relationship of ACM to the Weapons Industry, which requires discussion. This will be a difficult conversation, but can it be avoided? It appears to me, however, that the question and discussion around including the military has distracted from the massive implications that the current draft has on any technological effect, and how it provides a license to justify for oneself unethical computing designs with no need to seek justification externally. 

The suggestion that computing should continue to justify its own consequences without recourse to legitimate processes of justification is a terrifying prospect. The strengthening of the Code cannot wait until ACM has clarified its position with respect to military computing, and the key weaknesses that were opened up by the changes to Principle 1.2 can be addressed independently.

After the Code is before the Code. I urge ACM to open a new revision process right now.

\bibliographystyle{alpha}
\bibliography{acmethics}

\end{document}